\documentclass[prb,preprint]{revtex4-1}   

\usepackage{graphicx}
\usepackage{amsmath}
\usepackage{amssymb}
\usepackage{mathrsfs}
\usepackage{amsfonts}
\usepackage{dsfont}
\usepackage{dcolumn}
\usepackage{bm}
\usepackage{booktabs}
\usepackage{textcomp}

\everymath{\displaystyle}

\newcommand{\di}{\mathrm{d}}

\newcommand{\br}{\bm{r}}
\newcommand{\be}{\bm{e}}

\newcommand{\bk}{\bm{k}}

\newcommand{\bp}{\bm{p}}

\newcommand{\bP}{\bm{P}}
\newcommand{\bPi}{\bm{\Pi}}

\newcommand{\bE}{\bm{E}}
\newcommand{\bB}{\bm{B}}

\newcommand{\bV}{\bm{V}}
\newcommand{\bA}{\bm{A}}

\providecommand{\abs}[1]{\lvert#1\rvert}

\begin{document}

\title{Near field of an oscillating electric dipole and cross-polarization of a collimated beam of light: two sides of the same coin}
\author{Andrea Aiello$^{1,2}$}
\email{andrea.aiello@mpl.mpg.de}
\author{Marco Ornigotti$^1$}

%
\affiliation{$^1$ Max Planck Institute for the Science of Light, G$\ddot{u}$nther-Scharowsky-Strasse 1/Bau24, 91058 Erlangen,
Germany}
\affiliation{$^2$Institute for Optics, Information and Photonics, University of Erlangen-Nuernberg, Staudtstrasse 7/B2, 91058 Erlangen, Germany}
\date{\today}

\begin{abstract}
We address the question of whether there exists a hidden relationship between the near-field distribution generated by an oscillating electric dipole and the so-called cross polarization of a collimated beam of light. We find that the answer is affirmative by showing that the complex field distributions occurring in both cases have a common physical origin: the requirement that the electromagnetic fields must be transverse.
\end{abstract}

\maketitle

\section{Introduction}
The goal of this paper is to illustrate the somewhat hidden connection between two seemingly different quantities: the  near-field terms in the expression of the electric field emitted by an oscillating electric dipole, and the cross-polarization contributions to the electric field of a propagating beam of light. The term ``cross-polarization" is used in literature to denote the appearance of a component of the electric field orthogonal to both the direction of the wave vector $\bm{k}_0$ and the field polarization $\bm{E}_0$ due to the nonplanar character of the field's wave front \cite{ref1,ref2,ref3,ref4}. In order to better understand this concept, one could think about performing a simple experiment: take two crossed polarizers (i.e., a set of two perfect polarizers oriented along orthogonal directions) and examine the field transmitted by this device. Clearly, if the impinging field is a plane wave, the transmitted field will always be zero. Less intuitively, for a beam with a finite transverse extent, the transmitted field will always be non zero, due to the fact that a beam of light has a non planar wave front that induces a cross-polarization term. Actually, cross-polarization is a very important feature, that has been extensively used in many branches of science like biology, geology, chemistry and material science, to name a few. Moreover, cross-polarization also enters in our everyday life, since it is at the basis of the modern 3D movie technology \cite{crossPol}.
 
This work is organized as follows: section II introduces the reader to the use of the Hertz potentials to calculate the electromagnetic field irradiated by arbitrary sources. In Sec. III  this technique is applied to the calculation of the field emitted by an oscillating electric dipole at rest in the origin of a given  Cartesian reference frame. Section IV constitutes the main body of this manuscript, in which we derive the formulas for the so-called cross polarization of a collimated beam of light. Finally, in Sec. V we compare the field distributions obtained in the two previous sections and show their common physical origin. 
\section{Maxwell equations and Hertz vector potentials}
In this section we summarize the theory of Hertz vector potentials which can be found, for example, in the classical books by Stratton \cite{Stratton}, Jackson \cite{jackson} and in a more didactic form in Ref. \cite{Essex1977}. Let us begin by writing the Maxwell equations for the electric $\bE(\br,t)$ and magnetic $\bB(\br,t)$ fields in vacuum in absence of charges and currents as
\begin{subequations}\label{Maxwell1}
\begin{align}
\bm{\nabla} \cdot \bE  = & \; 0, \label{Maxwell1A} \\
\bm{\nabla} \cdot \bB  = & \; 0, \label{Maxwell1B} 
\end{align}
\end{subequations}
and 
\begin{subequations}\label{Maxwell2}
\begin{align}
\bm{\nabla} \times \bE  + \frac{\partial \bB}{\partial t} = & \; 0, \label{Maxwell2A} \\
\bm{\nabla} \times \bB  -\frac{1}{c^2} \frac{\partial \bE}{\partial t}   = & \; 0, \label{Maxwell2B} 
\end{align}
\end{subequations}
where $c$ is the speed of light in vacuum. By using the well-known formulas \cite{jackson}
\begin{subequations}\label{VecFor}
\begin{align}
\bm{\nabla} \times \left( \bm{\nabla} \psi \right) = & \; 0, \label{VecForA} \\
\bm{\nabla} \times   \left( \bm{\nabla} \times \bA \right)   = & \; \bm{\nabla} \left( \bm{\nabla} \cdot \bA \right) - \nabla^2 \bA, \label{VecForB} \\
\bm{\nabla} \cdot   \left( \bm{\nabla} \times \bA \right)   = & \; 0, \label{VecForC}%
\end{align}
\end{subequations}
which are valid for any scalar  and vector fields $\psi(\br)$ and $\bA(\br)$, respectively, 
it is possible to prove that the electric and magnetic fields written via the so-called \emph{Hertz vector potential} $\bm{\Pi}(\br,t)$ as 
\begin{subequations}\label{HertzFields}
\begin{align}
\bE(\br,t) = & \;  \bm{\nabla} \left( \bm{\nabla} \cdot \bm{\Pi} \right) - \frac{1}{c^2} \frac{\partial^2 \bm{\Pi}}{\partial t^2}, \label{HertzFieldsA} \\
\bB(\br,t) = & \; \frac{1}{c^2} \left( \bm{\nabla} \times   \frac{\partial \bm{\Pi}}{\partial t}  \right),
\label{HertzFieldsB}
\end{align}
\end{subequations}
identically satisfy  the Maxwell equations \eqref{Maxwell1} and \eqref{Maxwell2} provided that $\bm{\Pi}(\br,t)$ is a solution of the Helmholtz wave equation 
\begin{align}\label{HelmScal}
\nabla^2 \bm{\Pi} -\frac{1}{c^2} \frac{\partial^2 \bm{\Pi}}{\partial t^2}=0.
\end{align}
Throughout this article we shall use a unique Hertz vector potential denoted with $\bm{\Pi}$. However, in the literature, usually two independent potentials are introduced: the Hertz electric and magnetic vector potentials $\bm{\Pi}_e$ and $\bm{\Pi}_m$, respectively. The difference between $\bm{\Pi}_e$ and $\bm{\Pi}_m$ resides in the sources of external electric and magnetic polarization densities $\bm{P}_\text{ext}$ and $\bm{M}_\text{ext}$, respectively, that generate them \cite{jackson}.  Anyhow, in vacuum $\bm{P}_\text{ext}=0=\bm{M}_\text{ext}$ and a single Hertz vector potential $\bm{\Pi}$ can be used. According to the convention adopted by Jackson, our $\bm{\Pi}$  coincides with $\bm{\Pi}_e$ in vacuum.

The validity of the relation \eqref{HertzFieldsA}, for example, may be shown by noticing that Eq. \eqref{Maxwell2A} is achieved by applying the operator $\bm{\nabla} \times$ to both sides of Eq. \eqref{HertzFieldsA}:
\begin{align}\label{Prove1}
\bm{\nabla} \times \bE = & \;  \bm{\nabla} \times \left[\bm{\nabla} \left( \bm{\nabla} \cdot \bm{\Pi} \right)\right] - \frac{1}{c^2} \bm{\nabla} \times \frac{\partial^2 \bm{\Pi}}{\partial t^2}, \nonumber \\
 = & \; - \frac{\partial }{\partial t} \left[ \frac{1}{c^2} \left( \bm{\nabla} \times \frac{\partial \bm{\Pi}}{\partial t} \right) \right],
\nonumber \\
 = & \; - \frac{\partial \bB}{\partial t} ,
\end{align}
where Eq. \eqref{VecForA} has been used with $\psi = \bm{\nabla} \cdot \bm{\Pi}$. Similarly, 
 by applying the operator $\bm{\nabla} \cdot$ to both sides of the same equation one obtains
\begin{align}\label{Prove2}
\bm{\nabla} \cdot \bE = & \;  \bm{\nabla} \cdot \left[\bm{\nabla} \left( \bm{\nabla} \cdot \bm{\Pi} \right)\right] - \frac{1}{c^2} \bm{\nabla} \cdot \frac{\partial^2 \bm{\Pi} }{\partial t^2}, \nonumber \\
 = & \; \bm{\nabla} \cdot \left( \nabla^2 \bm{\Pi} -\frac{1}{c^2} \frac{\partial^2 \bm{\Pi}}{\partial t^2} \right) ,
\nonumber \\
 = & \;0 ,
\end{align}
where Eqs. \eqref{VecForB}, \eqref{VecForC} and \eqref{HelmScal} have been used. Verification of Eqs. \eqref{Maxwell1B} and \eqref{Maxwell2B}  is left to the reader as an exercise. It should be noticed that if we assume \emph{a priori} that Eq. \eqref{HelmScal} is satisfied by $\bm{\Pi}(\br,t)$, then we can rewrite Eqs. \eqref{HertzFields} in the simpler form
\begin{subequations}\label{HertzFields2}
\begin{align}
\bE(\br,t) = & \; \bm{\nabla} \times \left(\bm{\nabla} \times   \bm{\Pi} \right), \label{HertzFields2A} \\
\bB(\br,t) = & \; \frac{1}{c^2} \left( \bm{\nabla} \times   \frac{\partial \bm{\Pi}}{\partial t}  \right).
\label{HertzFields2B}
\end{align}
\end{subequations}
The latter equations are written in the form that we shall use in the remainder.
\section{Electric Dipole Fields}
\subsection{Hertz vector potential of an oscillating dipole}
In this section we shortly review the problem of calculating the exact electric and magnetic fields from an electric dipole with harmonic time dependence. Let $\bp \exp(- i \omega t)$ be the \emph{electric dipole moment} of a localized system of charges (source) that vary sinusoidally in time at angular frequency $\omega$. Without loss of generality, we assume that the source is centered at the origin 
of the Cartesian reference frame $\{x,y,z \}$ and
we choose the axis $z$  parallel to the dipole moment, namely $\bp = p \, \be_z$ with $p= \abs{\bp}$.
Then, the electric dipole fields can be calculated at any point $\br \neq \bm{0}$ 
from Eqs. \eqref{HertzFields2} by choosing \cite{jackson}
\begin{align}
\bm{\Pi}(\br,t) =  \bp  \, \frac{1}{4 \pi \epsilon_0} \frac{\exp \left[ i \left( k \, r -  \omega \, t \right)\right]}{r}, \label{Hertz}
\end{align}
where $k = \omega/c$, $\epsilon_0$ is the permittivity of vacuum and  $r = \abs{\br} = (x^2 + y^2 + z^2)^{1/2}$ is the distance from the origin of coordinates  to the point $\br$ along the radial direction $\be_r= \br/r$.
It should be noticed that  $\bm{\Pi}(\br,t)$ is equal to the product of the constant vector $\bp$ with the outgoing spherical wave $\exp \left[ i \left( k \, r -  \omega \, t \right)\right]/(4 \pi \epsilon_0 \, r)$ and, therefore, is a solution of the \emph{monochromatic} Helmholtz wave equation 
\begin{align}
{\nabla}^2 \bm{\Pi} + k^2 \bm{\Pi} =0. \label{Helmholtz}
\end{align}
We shall make use of this observation later.

 If we substitute Eq. \eqref{Hertz} into Eqs. \eqref{HertzFields2}, after some algebra we obtain the well known results, \cite{Farina2011}
\begin{widetext} 
\begin{subequations}\label{DipoleFields}
\begin{align}
\bE(\br,t) = & \; \frac{e^{-i \omega t}}{4 \pi \epsilon_0} \left\{ 
k^2 \left( \be_r \times \bp \right) \times \be_r \frac{e^{i k r}}{r} + 
\bigl[ 
 3 \be_r \left( \be_r \cdot \bp\right) - \bp
\bigr] \left( \frac{1}{r^3} -\frac{i k}{r^2}\right)e^{i k r}
\right\}, \label{DipoleFieldsA} \\
c \bB(\br,t) = & \; \frac{e^{-i \omega t}}{4 \pi \epsilon_0} \left\{ 
k^2 \left( \be_r \times \bp \right) \frac{e^{i k r}}{r} \left( 1 -\frac{1}{i k r}\right)
\right\},
\label{DipoleFieldsB}
\end{align}
\end{subequations}
\end{widetext}
where $\bm{e}_r=\bm{r}/r_0$. Equation \eqref{DipoleFieldsA} shows that $\bE(\br,t)$ is expressible as a sum of three terms in $1/r^3$, $1/r^2$ and $1/r$ which dominate in the near (static), intermediate (induction) and far (radiation) zone, respectively \cite{jackson2}. A thorough discussion of the physical meaning of these terms may be found, e.g., in Ref. \cite{jackson} and will not be repeated here. However,
 for the sake of completeness, we notice that in the radiation zone, where the terms in $1/r^3$ and $1/r^2$ can be neglected, it is natural to represent the fields using spherical coordinates $\{r, \theta, \phi \}$ as follows: 
\begin{subequations}\label{DipoleFieldsRad2}
\begin{align}
\bE^\text{rad}(\br,t) = & \; \mathscr{E}_0 
\,\be_\theta   \sin \theta \,\frac{\exp \left[ i \left( k \, r -  \omega \, t \right)\right]}{k r}
, \label{DipoleFieldsRad2A} \\
c\bB^\text{rad}(\br,t)  = & \; \mathscr{E}_0 
\,\be_\phi   \sin \theta \, \frac{\exp \left[ i \left( k \, r -  \omega \, t \right)\right]}{kr} ,
\label{DipoleFieldsRad2B}
\end{align}
\end{subequations}
 where $\mathscr{E}_0= - k^3 p/(4 \pi \epsilon_0)$ is a constant  with the dimensions of an electric field. Equations \eqref{DipoleFieldsRad2} show that upon the surface of the sphere of radius $r\gg 1/k$ centered at the dipole position $\br = \bm{0}$,  the electric field vector $\bE^\text{rad} \parallel \be_\theta$ and the magnetic field vector $\bB^\text{rad} \parallel \be_\phi$ are orthogonal each other and  perpendicular to the position vector $\br = r \be_r$.  In other words, in the radiation zone the dipole fields  behave, close to the point $\br$, as the fields of  a plane wave of wave vector $k \br/r$.
 Moreover, both fields vanish along the dipole axis $z$ where $\sin \theta = 0$. Last but not least, the fields \eqref{DipoleFieldsRad2} are \emph{not} exact solutions of neither Helmholtz  nor Maxwell equations.
 
 \subsection{Electromagnetic field generated by a monochromatic Hertz vector potential}
We conclude this section by rewriting Eqs.  \eqref{HertzFields} for the monochromatic Hertz vector potential Eq. \eqref{Hertz}  in a form that will be useful later for the comparison with the cross-polarization terms of a polarized beam of light. 
By using  Eq. \eqref{HelmScal} into  Eq. \eqref{HertzFieldsA} and noticing that from  Eq. \eqref{Hertz} it follows that  $-\nabla^2 \bm{\Pi} = k^2 \bm{\Pi}$ and $\partial \bm{\Pi} / \partial t = - i k c \bm{\Pi}$, we can rewrite the electromagnetic fields in Eqs. \eqref{HertzFields} as
\begin{subequations}\label{DipoleFields3}
\begin{align}
\bE(\br,t) = & \; k^2 \bm{\Pi} + \bm{\nabla} \left( \bm{\nabla} \cdot \bm{\Pi} \right), \label{DipoleFields3A} \\
c \bB(\br,t) = & \,  - i k  \bm{\nabla} \times  \bm{\Pi}  . \label{DipoleFields3B}
\end{align}
\end{subequations}
The physical interpretation of the result \eqref{DipoleFields3A} is straightforward: The Hertz vector potential $\bm{\Pi}(\br,t)$ (times $k^2$) can be regarded itself as an electric field which satisfies the Helmholtz wave equation Eq. \eqref{HelmScal} but not  Maxwell equation \eqref{Maxwell1A} because, in general, $k^2\bm{\nabla} \cdot \bm{\Pi} \neq 0$. Therefore, in order to obtain a bona fide electric field  one must add to the first terms  $k^2 \bm{\Pi}$ in Eq. \eqref{DipoleFields3A}, the second term $\bm{\nabla} \left( \bm{\nabla} \cdot \bm{\Pi} \right)$ whose divergence is equal to $\bm{\nabla} \cdot [\bm{\nabla} \left( \bm{\nabla} \cdot \bm{\Pi} \right)] = \nabla^2  \left( \bm{\nabla} \cdot \bm{\Pi} \right) = 
 \bm{\nabla} \cdot  \left( \nabla^2 \bm{\Pi} \right) = - k^2  \bm{\nabla} \cdot \bm{\Pi}$ which exactly cancels the divergence of the first term $\bm{\nabla} \cdot (k^2 \bm{\Pi})$ thus yielding to the correct result $\bm{\nabla} \cdot \bE=0$.
\section{Cross polarization}
In this section we summarize the main results of the study of  the polarization of nonplanar wave fronts \cite{FandS}. We will show that  the polarization of a plane wave field of wave vector $\bk_0$ can be completely specified by a single vector $\bE_0$ perpendicular to  $\bk_0$. Conversely, for a wave field with a nonplanar wave front the sole vector $\bE_0$ cannot suffice and electric field components parallel to  $\bk_0$ (longitudinal) and perpendicular to both  $\bE_0$  and $\bk_0$ (cross polarization) will appear as unavoidable consequence of the angular spread of the field.
\subsection{Plane wave vector fields}
%
%
 To begin with, consider an electromagnetic plane wave of angular frequency $\omega_0$ and wave vector $\bk_0$, with $\abs{\bk_0} = \omega_0/c$. Following Ref. \cite{jackson}, we write the plane wave fields, which are solutions of both Helmholtz  and Maxwell equations in vacuum, as 
\begin{subequations}\label{PWFields10}
\begin{align}
\bE(\br,t) = & \; \bE_0 \, \exp \left[ i \left(  \bk_0 \cdot \br -  \omega_0 t \right) \right]
, \label{PWFields10A} \\
\bB(\br,t)  = & \; \bB_0 \, \exp \left[ i \left(  \bk_0 \cdot \br -  \omega_0 t \right) \right] ,
\label{PWFields10B}
\end{align}
\end{subequations}
where $\bE_0 $  and $\bB_0 $ are constant (generally complex-valued) vectors such that
\begin{align}\label{ConstVec}
\bk_0 \cdot \bE_0  = 0  = \bk_0 \cdot \bB_0 ,
\end{align}
with 
\begin{align}\label{ConstVec2}
c \, \bB_0  = \frac{ \bk_0}{k_0}  \times \bE_0 .
\end{align}
It is a simple exercise to verify that Eq. \eqref{ConstVec} guarantees the validity of the divergence Maxwell   equations \eqref{Maxwell1}
while Eq.  \eqref{ConstVec2} ensures that $\bE$ and $\bB$ are solutions of Eqs. \eqref{Maxwell2}.
Thus, the polarization of a plane wave field can be represented by a \emph{single} vector $\bE_0$ perpendicular to the direction of wave propagation $\bk_0$. This observation lies at the foundations of the Jones and Mueller matrix calculi \cite{Hecht, MandelBook}.
In many practical applications beams of light are well collimated and their wave fronts are almost planar. In these cases the Jones and Mueller calculi are still applicable. 
\subsection{Nonplanar wave front scalar fields}
Consider now the case of an arbitrary wave field with (possibly) nonplanar wave front. With the locution ``wave field'' henceforth we indicate any either scalar or vector field which is an exact solution of the 
 Helmholtz wave equation Eq.  \eqref{HelmScal}.  Thus, let  $\psi(\br, t)$ be  a generic \emph{scalar} wave field and let denote with  $\widetilde{\psi} (\bk,t)$  the three-dimensional Fourier transform of $\psi(\br, t)$ calculated as
\begin{align}\label{HelmScalFT}
 \widetilde{\psi} (\bk,t)= \frac{1}{(2 \pi)^{3/2}} \int \psi(\br, t) \exp\left( -i \bk \cdot \br \right) \di^3 r.
\end{align}
Equation \eqref{HelmScal} requires  $\widetilde{\psi} (\bk,t)$ to satisfy the equation of motion 
\begin{align}\label{EqOfMot}
 \omega^2  \widetilde{\psi} + \frac{\partial^2 \widetilde{\psi}}{\partial t^2}=0 ,
\end{align}
with $\omega = c \abs{\bk}$. While for the fields \eqref{PWFields10} we had a single direction of wave propagation along the wave vector  $\bk_0$, in the present case there are many different wave vectors $\bk$ associated to the field $\psi(\br,t)$. However, we can still assign a principal (or, central) direction of propagation to $\psi(\br,t)$ by calculating the mean value of the wave vector $\bk$ with respect to the wave vector non-negative distribution $\abs{\widetilde{\psi} (\bk,t)}^2$:
\begin{align}\label{MeanK}
\overline{\bk} = \frac{\displaystyle{\int \bk \, \abs{\widetilde{\psi} (\bk,t)}^2  \di^3 k}}{\displaystyle{
\int \abs{\widetilde{\psi} (\bk,t)}^2  \di^3 k}} \equiv \bk_0 \, ,
\end{align}
where the subscript ``$0$'' here denotes the central wave vector \cite{note}.
Without loss of generality, we choose the Cartesian reference frame $\{x,y,z \}$ with the axis $z$ parallel to $\bk_0$, namely $\bk_0 = \abs{\bk_0} \be_z \equiv k_0 \, \be_z$. In this frame by definition one has $\overline{k_x}=0=\overline{k_y}$ and $\overline{k_z}=k_0$. The information about the deviation from planarity of the field wave front may be conveniently extracted from the covariance matrix $C$ \cite{Porras}  of the wave vector $\bk$ whose elements are $C_{ij} = \overline{k_i k_j} - \overline{k_i} \; \overline{k_j}$, where the $\overline{k_i}$s are defined via Eq. \eqref{MeanK} and
\begin{align}\label{VarK}
\overline{k_i k_j} = \frac{\displaystyle{\int k_i k_j \, \abs{\widetilde{\psi} (\bk,t)}^2  \di^3 k}}{\displaystyle{
\int \abs{\widetilde{\psi} (\bk,t)}^2  \di^3 k}}, \qquad i,j \in \{x,y,z \}  . 
\end{align}
It is not difficult to show (see Appendix A) that for a plane wave field the covariance matrix $C$ is identically zero while for a spherical wave $C = (4 \pi/3)I_3$, where $I_3$ is the $3\times 3$ identity matrix (as expected for an isotropic distribution).
More generally, for any wave field,  the \emph{angular spread} $\theta_0$ around the direction of propagation $z$ can be defined as:
\begin{align}\label{AngSpread}
\theta_0 \equiv \frac{2^{1/2}}{k_0} \left(C_{xx} +C_{yy} \right)^{1/2}  ,
\end{align}
where $C_{xx} +C_{yy}$ is the variance of the transverse components, with respect to the $z$-axis of propagation, of the wave vector $\bk$.
This (arbitrary) choice for such a measure is physically motivated by the fact that in the limiting case of a Gaussian monochromatic paraxial field of waist $w_0$ and  Rayleigh range  $L = k_0 w_0^2/2$, 
\begin{align}\label{ParField}
\psi(\br,t) =  \frac{\exp \left[ i k_0 \left(z - c t \right) \right]}{1+iz/L} \exp \left[ - \frac{1}{w_0^2} \left( \frac{x^2 + y^2}{1+iz/L} \right) \right], 
\end{align}
equations \eqref{VarK}  and \eqref{AngSpread} give (see Appendix B for details of the calculation):
\begin{align}\label{ParField2}
\theta_0 = \frac{2}{k_0 w_0} \qquad \text{and}\qquad \overline{k_z^2}/k_0^2 = \cos^2\theta_0 + O( \theta_0^4), 
\end{align}
in agreement with the standard theory of paraxial beams \cite{Siegman}.
\subsection{Nonplanar wave front vector fields}
The considerations above are valid for arbitrary \emph{scalar} wave fields, however, in this work we want to study some properties of realistic electromagnetic \emph{vector} fields. Therefore, we need to understand how to attach a polarization vector to a scalar wave field in order to achieve  bona fide  electric and magnetic fields.

In most actual experimental situations a linearly polarized collimated beam of light is prepared by inserting a polarizing plate (e.g., a polaroid sheet) in a plane perpendicular to the direction of propagation $z$ of the beam. In this case, the transmission axis of the linear polarizer can be represented by a real-valued transverse unit vector, say $\bP_0 = P_{0x} \be_x + P_{0y} \be_y$, with $P_{0x}^2 + P_{0y}^2=1$. Similarly, circular polarization may be achieved by placing  a  quarter-wave plate (retarder) after the linear polarizer \cite{Hecht}. For such a composite optical system (wave plate plus polarizer) the transmission axis will be still described by the vector $\bP_0$ which, however, now  takes complex values and the normalization conditions reads as $\abs{P_{0x}}^2 + \abs{P_{0y}}^2=1$. 
Thus,  it seems a reasonable assumption to write the electric field of a well collimated beam of light in the form of a product between a constant polarization vector $\bP_0$ and a wave field $\psi(\br,t)$:
\begin{align} \label{PWFields30} 
\bE(\br,t) = & \;  \bP_0 \, \psi(\br,t).
\end{align}
However, this assumption is plagued by a serious problem when the beam is not collimated. In fact, in this case   the wave front is not planar and, therefore, Eq. \eqref{Maxwell1A} cannot be satisfied:
\begin{align}
\bm{\nabla} \cdot \bE =   \bP_0 \cdot \left( \bm{\nabla} \psi\right) \neq  0.  \label{PWFields40} 
\end{align}
In fact,  one would need a plane wave field of the form $\psi(\br,t) = (2 \pi)^{-3/2}\exp[i(k_0 z - \omega_0 t)]$ to achieve $\bm{\nabla} \psi = i k_0 \psi \,\be_z \perp \bP_0$. The origin of this problem resides in the fact that while $\bP_0$ is a unique constant vector, there are infinitely many $\bk$ vectors in the wave field $\psi(\br,t)$ that are not perpendicular to $\bP_0$. This can be easily seen by rewriting Eq. \eqref{PWFields40}  in terms of the 3D Fourier transform of $\psi(\br,t)$ as
\begin{align}\label{PWFields50} 
\bm{\nabla} \cdot \bE = & \;  \bP_0 \cdot \left( \bm{\nabla} \psi\right) \nonumber \\
  = & \;  \frac{i}{(2 \pi)^{3/2}} \int \left( \bP_0 \cdot \bk \right) \widetilde{\psi}(\bk, t) \exp\left( i \bk \cdot \br \right) \di^3 k ,
\end{align}
where $\bP_0 \cdot \bk \neq 0$ for all $\bk \neq \bk_0 = k_0 \, \be_z$. However, for any given wave vector $\bk$ one can always decompose the vector $\bP_0$ in a transverse part and in a longitudinal part with respect to $\bk$. This can be proved by using the vector identity \cite{jackson}
\begin{align}\label{VecId} 
\bk \times(\bk \times \bA) = \bk (\bk \cdot \bA) - k^2 \bA,
\end{align}
from which it follows that
\begin{align}\label{PWFields60} 
\bP_0 = & \;   \frac{1}{k^2} \bigl[ \bk \left(\bk \cdot \bP_0 \right) - \bk \times \left(\bk \times \bP_0 \right) \bigr] \nonumber \\
\equiv & \;  \bP_\parallel + \bP_\perp,
\end{align}
where, by definition, the vector $\bP_\perp(\bk) =  - \bk \times \left(\bk \times \bP_0 \right)/k^2$ is purely transverse, namely $\bk \cdot \bP_\perp =0$, and the vector $\bP_\parallel(\bk)=\bk \left(\bk \cdot \bP_0 \right)/k^2$ is completely longitudinal, namely $\bk \times \bP_\parallel =0$. 
By using Eq. \eqref{PWFields60} and the Fourier transform of $\psi(\br,t)$ we can rewrite Eq. \eqref{PWFields30} as
\begin{align}\label{PWFields70} 
\bE(\br,t) = & \;  \bP_0 \, \psi(\br,t) \nonumber \\
= & \; \frac{1}{(2 \pi)^{3/2}} \int  \bigl( \bP_\parallel +  \bP_\perp \bigr)\widetilde{\psi}(\bk, t) \exp\left( i \bk \cdot \br \right) \di^3 k \nonumber \\
\equiv & \; \bE_\parallel(\br,t)+\bE_\perp(\br,t),
\end{align}
where, by definition, 
\begin{align}\label{PWFields80} 
\bm{\nabla} \cdot \bE_\perp =0, \qquad \text{and} \qquad \bm{\nabla} \times \bE_\parallel =0.
\end{align}
This result is a particular case of the more general Helmholtz theorem which states that an arbitrary  vector field $\bV(\br)$ in three-dimensional space can be  decomposed uniquely into a transverse part, $\bV_\perp$, and a longitudinal one, $\bV_\parallel$, in such a way that $\bm{\nabla} \cdot \bV_\perp =0= \bm{\nabla} \times \bV_\parallel$ \cite{Rohrlich2004,CTan}.
From the first of Eqs. \eqref{PWFields80}  it follows that $\bE_\perp(\br,t)$ is a good candidate for a realistic electric vector field. It is convenient for later purposes to write it explicitly in the Fourier transform form as
\begin{widetext} 
\begin{align}\label{PWFields90} 
\bE_\perp(\br,t) = & \; \frac{1}{(2 \pi)^{3/2}} \int  - \bk \times \bigg[\bk \times \bigg( \frac{\bP_0 \, \widetilde{\psi}(\bk, t)}{k^2}  \bigg) \bigg]  \exp\left( i \bk \cdot \br \right) \di^3 k \nonumber \\
 = & \; \frac{1}{(2 \pi)^{3/2}} \int  - \bk \times \Big[\bk \times \widetilde{\bP}(\bk,t)  \Big] \exp\left( i \bk \cdot \br \right) \di^3 k \nonumber \\
= & \; \frac{1}{(2 \pi)^{3/2}} \int   \widetilde{\bE}_\perp(\bk, t)  \exp\left( i \bk \cdot \br \right) \di^3 k,
\end{align}
\end{widetext} 
where we have defined $\bP_0 \widetilde{\psi}(\bk, t)/k^2 \equiv \widetilde{\bP}(\bk,t)$ and the last equality  simply defines the Fourier transform $\widetilde{\bE}_\perp(\bk, t)$ of the field. It should be noticed  that although Eq. \eqref{PWFields90} has been derived for a specific choice for $\widetilde{\bP}(\bk,t)$, it holds also for more general cases. In fact,
since $ \widetilde{\psi}(\bk,t)/k^2$ satisfies the same equation of motion \eqref{EqOfMot} as $\widetilde{\psi}(\bk,t)$ does and $\bP_0$ is time-independent, then the same equation is satisfied by $\widetilde{\bP}(\bk,t)$. Therefore, from the equation above we can write
\begin{align}\label{Etilde}
\widetilde{\bE}_\perp(\bk, t) = - \bk \times \left[\bk \times \widetilde{\bP}(\bk,t)  \right], 
\end{align}
where from now on  $\widetilde{\bP}(\bk,t)$ may be understood as  the Fourier transform of \emph{any} arbitrary vector solution $\bP(\br,t)$ of the Helmholtz wave equation:
\begin{align}\label{HelmScal2}
\nabla^2 \bP -\frac{1}{c^2} \frac{\partial^2 \bP}{\partial t^2}=0.
\end{align}

The  magnetic field $\bB(\br,t)$ generated by $\bP(\br,t)$ may be determined from  $\widetilde{\bE}_\perp(\bk, t)$ by noticing that, as shown in Appendix \ref{B}, the Maxwell equations \eqref{Maxwell1} and \eqref{Maxwell2} written in terms of the Fourier transform $\widetilde{\bE}_\perp(\bk, t)$ and $\widetilde{\bB}(\bk, t)$ of the electric and magnetic fields $\bE_\perp(\br,t)$ and $\bB(\br,t)$, respectively, read as \cite{CTan}
\begin{subequations}\label{MaxwellFT1}
\begin{align}
i \bk \cdot \widetilde{\bE}_\perp  = & \; 0, \label{MaxwellFT1A} \\
i \bk \cdot  \widetilde{\bB}  = & \; 0, \label{MaxwellFT1B} \\
i \bk \times  \widetilde{\bE}_\perp = & \; - \frac{\partial \widetilde{\bB}}{\partial t} , \label{MaxwellFT2A} \\
i \bk \times  \widetilde{\bB}     = & \; \frac{1}{c^2} \frac{\partial \widetilde{\bE}_\perp}{\partial t}. \label{MaxwellFT2B} 
\end{align}
\end{subequations}
By multiplying the left side of Eq. \eqref{MaxwellFT2B}  by $\bk \times$ and using Eq. \eqref{VecId} we obtain 
\begin{align}\label{PWFields100}
\bk \times \bigl( i \bk \times  \widetilde{\bB} \bigr)    = & \; \bk (i \bk \cdot \widetilde{\bB}) - i k^2 \widetilde{\bB} \nonumber \\
   = & \; - i k^2 \widetilde{\bB},
\end{align}
where  Eq. \eqref{MaxwellFT1B} has been used. Analogously, if we multiply the right side of Eq. \eqref{MaxwellFT2B}  by $\bk \times$ and use again Eq. \eqref{VecId} together with the definition Eq. \eqref{Etilde} we obtain
\begin{align}\label{PWFields110}
\bk \times \left( \frac{1}{c^2} \frac{\partial \widetilde{\bE}_\perp}{\partial t} \right)    = & \;\frac{1}{c^2} \frac{\partial }{\partial t}  \left( \bk \times \widetilde{\bE}_\perp \right)  \nonumber \\
   = & \; \frac{1}{c^2} \frac{\partial }{\partial t}  \left( \bk \times \left\{ - \bk \times 
   \left[ \bk \times \widetilde{\bP}(\bk,t)  \right] \right\} \right),  \nonumber \\
   = & \; \frac{k^2}{c^2} \bk \times \frac{\partial }{\partial t}   \widetilde{\bP}(\bk,t) .
\end{align}
Finally, by equating the right sides of Eqs. \eqref{PWFields100} and \eqref{PWFields110} we find the sought expression for $\widetilde{\bB}(\bk,t)$:
\begin{align}
\widetilde{\bB}(\bk,t) = & \; \frac{1}{c^2} \left[ i \bk \times\frac{\partial }{\partial t}   \widetilde{\bP}(\bk,t)  \right] . \label{Btilde} 
\end{align}
At this point, we can recollect our main results Eq. \eqref{Etilde} and Eq. \eqref{Btilde} and write
\begin{subequations}\label{PWFields130}
\begin{align}
\widetilde{\bE}_\perp(\bk,t) =  & \;  - \bk \times \left[\bk \times \widetilde{\bP}(\bk,t)  \right] , \label{PWFields130A} \\
\widetilde{\bB}(\bk,t) = & \; \frac{1}{c^2} \left[ i \bk \times\frac{\partial }{\partial t}   \widetilde{\bP}(\bk,t)  \right] . \label{PWFields130B} 
\end{align}
\end{subequations}

From Eq. \eqref{PWFields130A} the origin of the longitudinal and cross-polarization terms in the expression of the electric field becomes manifest when using Eq. \eqref{VecId} to rewrite $\widetilde{\bE}_\perp(\bk,t)$ in the form
\begin{align}\label{PWFields140}
\widetilde{\bE}_\perp(\bk,t) =  & \;  k^2 \widetilde{\bP} - \frac{1}{k^2} \bk \bigl( \bk \cdot \widetilde{\bP} \bigr) , \nonumber \\
 =  & \;  \left[ \bP_0  - \frac{1}{k^2} \bk \bigl( \bk \cdot \bP_0 \bigr)\right] \widetilde{\psi}(\bk,t), \nonumber \\
 =  & \;  \widetilde{\bE}(\bk,t) - \frac{1}{k^2} \bk \bigl[ \bk \cdot \widetilde{\bE}(\bk,t) \bigr] , 
\end{align}
where the definition  $\widetilde{\bP}(\bk,t) = \bP_0 \widetilde{\psi}(\bk,t)/k^2$ and the Fourier transform of Eq. \eqref{PWFields30} have been used.
The first term in Eq. \eqref{PWFields140} simply yields the original field $\bE(\br,t)$ given in 
 Eq. \eqref{PWFields30} which is not an exact solution of Maxwell equations. The second term is responsible for: \emph{a}) quadratic corrections to the main polarization $\bP_0$, \emph{b}) cross-polarization terms, and \emph{c}) longitudinal component of the field \cite{ref4}. This can be seen in a clearer manner by choosing, for example, a field linearly polarized along the axis $x$ for which one has $\bP_0 = \be_x$. Then, a straightforward calculation furnishes
\begin{widetext} 
\begin{align}\label{PWFields150}
\bP_0- \frac{1}{k^2}\bk \bigl( \bk \cdot \bP_0 \bigr)  =  & \; \be_x- \frac{1}{k^2}\biggl(
\underbrace{ k_x^2 \, \be_x  }_{\text{quadratic}}
+ \underbrace{k_x k_y \,  \be_y }_\text{cross-polarization} 
+ \underbrace{k_z  k_x  \, \be_z}_\text{longitudinal} \biggr). 
\end{align}
\end{widetext} 
According to our previous discussion about how to prepare a beam of light polarized along the direction $\bP_0$, Eq. 
\eqref{PWFields150} shows that even when one sets the polarizer oriented along the direction $\be_x$ (first term in the right side of Eq. 
\eqref{PWFields150}), the beam resulting from transmission  will unavoidably  posses components both along the transverse direction $\be_y$ and the direction of propagation $\be_z$.
\section{Cross-polarization versus near field}
What is the connection between the beam fields given by Eqs. \eqref{PWFields130} and the electric dipole fields  Eqs. \eqref{DipoleFields}? In order to answer this question we must first Fourier transform backwards Eqs. \eqref{PWFields130} to obtain the configuration space fields $\bE_\perp(\br,t)$ and $\bB(\br,t)$. To this end, let us multiply both sides of Eqs. \eqref{PWFields130} by $(2 \pi)^{-3/2} \exp(i \bk \cdot \br)$ and integrate with respect to $\di^3 k = \di k_x \di k_y \di k_z$ to obtain
\begin{subequations}\label{PWFields160}
\begin{align}
\bE_\perp(\br,t) =  & \;\frac{-1}{(2 \pi)^{3/2}} \int  \bk \times \left[\bk \times \widetilde{\bP}(\bk,t)  \right] \exp(i \bk \cdot \br) \di^3 k , \label{PWFields160A} \\
\bB(\br,t) = & \; \frac{1}{(2 \pi)^{3/2}c^2} \int  i \bk \times\frac{\partial }{\partial t}   \widetilde{\bP}(\bk,t)   \exp(i \bk \cdot \br) \di^3 k. \label{PWFields160B} 
\end{align}
\end{subequations}
For a vector field of the form  $\bA(\br) = \bA_0 \exp(i \bk \cdot \br)$, being $\bA_0$ a constant vector, a straightforward calculation furnishes 
\begin{subequations}\label{PWFields170}
\begin{align}
\bm{\nabla} \times \bA  =  & \; i \bk \times \bA , \label{PWFields170A} \\
\bm{\nabla} \times \bigl( \bm{\nabla} \times \bA \bigr) =  & \; - \bk \times \bigl( \bk \times \bA \bigr). \label{PWFields170B} 
\end{align}
\end{subequations}
By using Eqs. \eqref{PWFields170} into Eqs. \eqref{PWFields160} one obtains the sought expressions for the fields of the beam:
\begin{subequations}\label{PWFields180}
\begin{align}
\bE_\perp(\br,t) =  & \; \bm{\nabla} \times \left(\bm{\nabla} \times   \bP \right), \label{PWFields180A} \\
\bB(\br,t) = & \;  \frac{1}{c^2} \left( \bm{\nabla} \times   \frac{\partial \bP}{\partial t}  \right), \label{PWFields180B} 
\end{align}
\end{subequations}
where $\bP(\br,t)$ denotes the Fourier transform of $\widetilde{\bP}(\bk,t) $.

Surprisingly enough, Eqs. \eqref{PWFields180} are identical to Eqs. \eqref{HertzFields2} if we identify $\bP(\br,t)$ with the Hertz vector potential $\bm{\Pi}(\br,t)$. 
Thus, the procedure to attach a polarization vector to a scalar wave field developed in the previous section, automatically leads to 
the introduction of the Hertz vector potential. This identification is consistent with the fact that both $\bPi$ and $\bP$ are solutions of the Helmholtz wave equations \eqref{HelmScal} and \eqref{HelmScal2}, respectively. This is also congruous with the observation made at the end of Sec. III where it was noticed that the Hertz potential could be regarded as an electric field itself. Moreover, both  $\bP$ and $\bPi$ are of the form of a product between a constant vector and a wave field $\psi(\br,t)$, the only difference resides in the form of  $\psi(\br,t)$. Here we can see why the near field terms of an oscillating electric dipole and the cross-polarization ones of a collimated beam of light are two sides of the same coin: in both cases these terms have a form which is determined by the vector operator $\bm{\nabla} \times \left(\bm{\nabla} \times   \circ \right)$ and they becomes negligible when the distance from the origin of coordinates increases. While this latter fact is well known for dipole fields, it is not appreciated enough for beam fields. Therefore, now we calculate explicitly Eqs. \eqref{PWFields180} for the monochromatic paraxial beam Eq. \eqref{ParField} that we rewrite in terms of the scaled coordinates $X = x/w_0$, $Y = y/w_0$ and $Q(z) = z/L -i$, as
\begin{align}\label{ParField2b}
\psi(\br,t) = E_0  \frac{\, \exp \left[ i k_0 \left(z - c \, t \right) \right]}{i \, Q(z)} \exp \left[ - \frac{X^2 + Y^2}{i \,Q(z)}  \right], 
\end{align}
where $E_0$ is a real constant with the dimensions of an electric field. For the sake of definiteness we choose $\bP_0 = \be_x$ and, according to the previous section, the ``Hertz vector potential'' $\bP(\br,t)$ will be chosen as
\begin{align}\label{ParField3}
\bP(\br,t) = \be_x \, \frac{\psi(\br,t)}{k_0^2}. 
\end{align}
By inserting Eq. \eqref{ParField3} into Eqs.  \eqref{PWFields180}, we obtain after a long but straightforward calculation the following expressions for the electromagnetic fields:

\begin{widetext}
\begin{subequations}\label{ParField4}
\begin{align}
\bE_\perp(\br,t) = & \; \psi(\br,t)  \Bigg\{ 
\be_x \left[
1+\theta_0^2\left(\frac{i}{2 \, Q}-\frac{X^2}{Q^2}\right) -\theta_0^4\left(\frac{1}{2 \, Q^2}+\frac{i \left(X^2+Y^2\right)}{Q^3}-\frac{\left(X^2+Y^2\right)^2}{4 \, Q^4}\right) \right]  \Bigg. \nonumber \\
& \; \phantom{\psi(\br,t)  \Bigg\{ } \Bigg. -\be_y  \left( \theta_0^2 \, \frac{X Y }{Q^2} \right) - \be_z \left[ \theta_0\frac{X }{Q}-\theta_0^3 \,\frac{X }{Q} \left(\frac{i }{Q}-\frac{X^2+Y^2}{2 \, Q^2}\right) \right] \Bigg\}, \label{ParField4A}\\
c \bB(\br,t) = & \; \psi(\br,t)  \Biggl\{\be_y \left[1+\frac{\theta_0^2}{2}\left(\frac{i}{ Q}-\frac{X^2+Y^2}{ Q^2}\right)\right]  - \be_z \, \theta_0 \frac{Y }{Q}\Biggr\}, \label{ParField4B}
\end{align}
\end{subequations}
\end{widetext}
where $\theta_0=2/(k_0 w_0) \ll 1$ is again the angular spread of the beam.  The latter plays the role of a small expansion parameter in the expressions above. Therefore, if we keep only the lowest order terms in $\theta_0$ for each polarization $\bm{e}_x$, $\bm{e}_y$  and $\bm{e}_z$, Eqs. \eqref{ParField4} reduce to the much more clearer following forms: 
\begin{subequations}\label{ParField5}
\begin{align}
\bE_\perp(\br,t) \simeq & \; \psi(\br,t)  \left[ 
\be_x  - \be_y \left(\theta_0^2 \, \frac{X Y }{Q^2} \right)  -  \be_z  \left(\theta_0 \, \frac{X }{Q} \right)\right], \label{ParField5A}\\
c \bB(\br,t)  \simeq & \; \psi(\br,t)  \left[ \be_y  -  \be_z \left(  \theta_0 \, \frac{Y }{Q} \right) \right] \label{ParField5B}.
\end{align}
\end{subequations}
By inspecting Eq. \eqref{ParField5A}, one can easily recognize both the well-known ``cloverleaf'' ($\be_y \, XY$ term) and the ``$\text{TEM}_{10}$'' ($\be_z \, X$ term) patterns of cross-polarization and longitudinal terms, respectively \cite{ref4}. The intensity patterns of the $x$-, $y$- and $z$-components of $\bE_\perp(\br,t)$ are depicted in Figs. \ref{figure1}, \ref{figure2} and \ref{figure3} respectively.

It should be noticed that from Eqs. \eqref{ParField5} it follows that $\bE_\perp \cdot \bB =0$, as for a plane wave field, but at the first order in $\theta_0$
\begin{align}
\text{Re} \left(\bE_\perp \times \bB^*\right) \propto & \; \text{Re} \left[\be_z + \theta_0 \left(  \be_x \, \frac{X }{Q}+ \be_y \, \frac{Y }{Q^*} \right) \right]\nonumber \\
= & \; \be_z + \theta_0 \,\frac{ Z}{1+Z^2}\left(  \be_x \, X+ \be_y \, Y \right),
\end{align}
where $Z = z/L$. This equation tells us that the cycle-averaged Poynting vector of the beam, proportional to $\text{Re} \left(\bE_\perp \times \bB^*\right) $, is not parallel to $\be_z$ but it takes a small \emph{radial} component when evaluated at a nonzero transverse distance from the axis $z$, as shown, e.g., in Fig.1 of Ref.  \cite{ref4}.

In the radiation zone where $z \gg L$, one has  $Q(z) \approx z/L \rightarrow \infty$ and the fields take on the limiting forms, 
\begin{subequations}\label{ParField6}
\begin{align}
\bE_\perp(\br,t) \simeq & \; \be_x \, \psi(\br,t) = k_0^2 \, \bP , \label{ParField6A}\\
c \bB(\br,t)  \simeq & \; \be_y\,  \psi(\br,t) = \be_z \times \bE_\perp, \label{ParField6B}
\end{align}
\end{subequations}
showing the characteristic plane wave behavior of the radiation fields (see analogous discussion after Eq. \eqref{DipoleFieldsRad2} for dipole fields).
\section{Concluding remarks}
In conclusions, we have shown that the dipole fields in the near and intermediate zones \cite{jackson2} have the same physical nature of the longitudinal and cross-polarization fields of a propagating beam. Both dipole and beam fields in the near region originates from the requirement that the electromagnetic fields must be transverse, namely that $\bm{\nabla} \cdot \bE = 0 = \bm{\nabla} \cdot \bB$. Moreover, they exhibit a plane wave-like behavior in the radiation zone.

%
\appendix{
\section{Covariance matrix for a plane wave vector field}\label{A}
Let 
\begin{align}\label{A10}
\psi(\br,t) = \frac{1}{(2 \pi)^{3/2}} \exp(i \bk_0 \cdot \br),
\end{align}
be a plane wave field. Its Fourier transform is, by definition, the three-dimensional Dirac delta function
\begin{align}\label{A20}
\widetilde{\psi}(\bk,t) = & \; \frac{1}{(2 \pi)^3} \int \exp\left[i \left(\bk_0 - \bk \right) \cdot \br\right] \di^3 k, \nonumber \\
= & \; \delta \left(\bk - \bk_0 \right)\nonumber \\
= & \; \delta \left(k_x -  k_{0x}\right)\delta \left(k_y -  k_{0y} \right)\delta \left(k_z -  k_{0z} \right).
\end{align}
In the reference frame where $\bk_0 = k_0 \be_z$ the expression above reduces to 
\begin{align}\label{A30}
\widetilde{\psi}(\bk,t) = \delta \left(k_x \right)\delta \left(k_y \right)\delta \left(k_z - k_0 \right).
\end{align}
In order to calculate the covariance matrix $C$ of the field \eqref{A10} let us evaluate first
\begin{align}\label{A40}
\overline{k_i^n k_j^m} = \frac{\displaystyle{\int k_i^n k_j^m \, \abs{\widetilde{\psi} (\bk,t)}^2  \di^3 k}}{\displaystyle{
\int \abs{\widetilde{\psi} (\bk,t)}^2  \di^3 k}},  
\end{align}
where $n,m$ are nonnegative integers and $i,j \in \{ x,y,z\}$. The wave vector distribution function $\abs{\widetilde{\psi} (\bk,t)}^2$ is a highly singular function involving powers of Dirac deltas. We overcome this difficulty by replacing one of the two delta functions $\delta(u)$ in each quadratic expression $\delta^2(u)$ by its finite limiting distribution $g_\epsilon(u)$ defined in such a way such that for any smooth function $f(u)$ one has \cite{DiracDelta}
\begin{align}\label{A50}
\lim_{\epsilon \to 0} \int g_\epsilon(u) f(u) \di u = f(0). 
\end{align}
 For example, $g_\epsilon(u)$ may be taken as a Gaussian distribution $g_\epsilon(u)= (\epsilon \, \pi^{1/2})^{-1} \exp(- u^2/\epsilon^2)$. 
Thus, in practice, we will make the following replacement every time is necessary:
\begin{align}\label{A60}
\delta^2(u) \rightarrow g_\epsilon(u) \delta(u)  = g_\epsilon(0) \delta(u), 
\end{align}
where now $g_\epsilon(0)$ is a finite quantity for $\epsilon \neq 0$. Eventually, at the end of our calculations, we can take the limit $\epsilon \to 0$:
\begin{align}\label{A70}
\overline{k_i^n k_j^m} = & \;\frac{\displaystyle{k_{0i}^n k_{0j}^m \, g_\epsilon^2(0) g_\epsilon(k_0) \! \int \! \delta \left(k_x \right)\delta \left(k_y \right)\delta \left(k_z - k_0 \right)  \di^3 k}}{\displaystyle{
g_\epsilon^2(0) g_\epsilon(k_0) \! \int \! \delta \left(k_x \right)\delta \left(k_y \right)\delta \left(k_z - k_0 \right)  \di^3 k}}  \nonumber \\
= & \;k_{0i}^n k_{0j}^m.
\end{align}
Since $k_{0x}=0=k_{0y}$ and $k_{0z} = k_0 = \overline{k_z}$ the result $C=0$ follows.
%
%
\begin{figure}[!t]
\begin{center}
\includegraphics[width=0.5\textwidth]{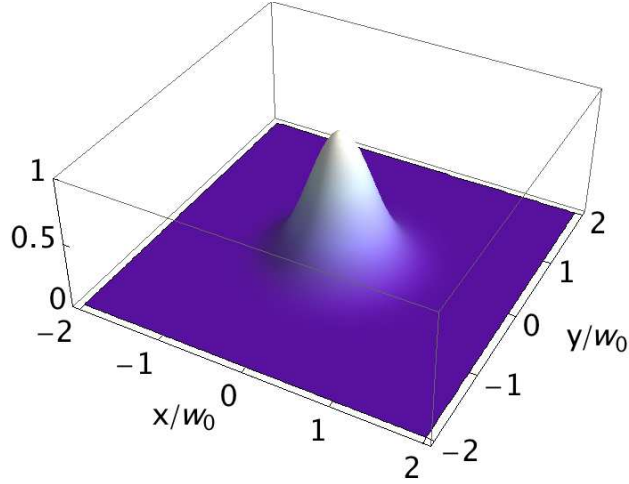}
\caption{Three dimensional plot of the normalized intensity $|\bm{E}_{\perp}\cdot\bm{e}_x|^2/E_0^2$ evaluated at $z=0$ of the $x$-component of the electric field as given by Eq. \eqref{ParField5A}. This constitutes the main contribution to the total intensity $|\bm{E}_{\perp}(\bm{r},t)|^2$ since, according to Eq. \eqref{ParField5A}, the intensities of the $y-$ and $z-$component are of order $\theta_0^4$ and $\theta_0^2$ respectively.}
\label{figure1}
\end{center}
\end{figure}
\begin{figure}[!t]
\begin{center}
\includegraphics[width=0.5\textwidth]{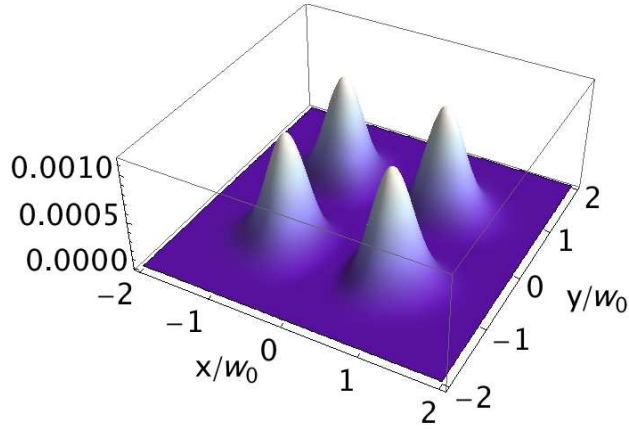}
\caption{Three dimensional plot of the normalized intensity $|\bm{E}_{\perp}\cdot\bm{e}_y|^2/(E_0^2\theta_0^4)$ evaluated at $z=0$ of the $y$-component of the electric field as given by Eq. \eqref{ParField5A}. The intensity pattern shows the characteristic ``cloverleaf" structure, i.e., four equal peaks all having the same distance from the origin of the axis $\{x,y\}=\{0,0\}$. This pattern is typical of cross-polarization.}
\label{figure2}
\end{center}
\end{figure}

\section{Maxwell's equations in Fourier space}\label{B}
In this Appendix we show how to calculate the Maxwell equations in the form given in Eqs. \eqref{MaxwellFT1}, for the Fourier transforms of the electric and magnetic fields defined via the following relations:
\begin{subequations}\label{FTfields}
\begin{eqnarray}
 \bE_\perp(\br, t)& = & \frac{1}{(2 \pi)^{3/2}} \int \! \! \widetilde{\bE}_\perp (\bk,t) \exp\left( i \bk \cdot \br \right) \di^3 k, \label{FTe} \\
\bB(\br, t) & = & \frac{1}{(2 \pi)^{3/2}} \int \! \! \widetilde{\bB} (\bk,t) \exp\left( i \bk \cdot \br \right) \di^3 k. \label{FTb}
\end{eqnarray}
\end{subequations}
Thus, for example, by substituting Eq. \eqref{FTe} into \eqref{Maxwell1A} one obtains
\begin{align}\label{B10}
\bm{\nabla} \cdot \bE_\perp= & \;\frac{1}{(2 \pi)^{3/2}} \int \bm{\nabla} \cdot  \left[\widetilde{\bE}_\perp (\bk,t) \exp\left( i \bk \cdot \br \right)\right] \di^3 k \nonumber \\
= & \;\frac{1}{(2 \pi)^{3/2}} \int \left[i \bk \cdot  \widetilde{\bE}_\perp (\bk,t)\right]  \exp\left( i \bk \cdot \br \right)\di^3 k,
\end{align}
from which Eq. \eqref{MaxwellFT1A} immediately follows after imposing $\bm{\nabla} \cdot \bE_\perp=0$. The same calculation can be easily repeated for $\bm{\nabla} \cdot \bB=0$.

Analogously, by substituting Eq. \eqref{FTe} into the expression $\bm{\nabla} \times \bE_\perp $ one finds
\begin{align}\label{B20}
\bm{\nabla} \times \bE_\perp = & \;\frac{1}{(2 \pi)^{3/2}} \int \bm{\nabla} \times  \left[\widetilde{\bE}_\perp (\bk,t) \exp\left( i \bk \cdot \br \right)\right] \di^3 k \nonumber \\
= & \;\frac{1}{(2 \pi)^{3/2}} \int \left[i \bk \times  \widetilde{\bE}_\perp (\bk,t)\right]  \exp\left( i \bk \cdot \br \right)\di^3 k.
\end{align}
Then, from Eqs. (\ref{Maxwell2A},\ref{B20}) and from the fact that 
\begin{align}\label{B30}
\frac{\partial \bB}{\partial t} = \frac{1}{(2 \pi)^{3/2}} \int \frac{\partial \widetilde{\bB} (\bk,t)}{\partial t}  \exp\left( i \bk \cdot \br \right)\di^3 k,
\end{align}
equation \eqref{MaxwellFT2A} straightforwardly follows. By following a  similar reasoning, Eq. \eqref{MaxwellFT2B} may be achieved.
%
%
\begin{figure}[!t]
\begin{center}
\includegraphics[width=0.5\textwidth]{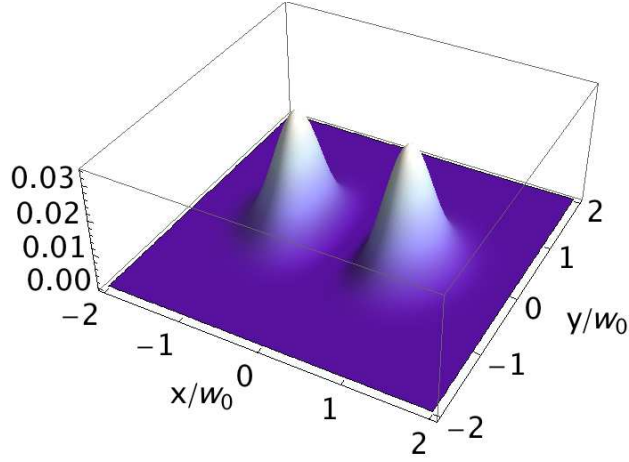}
\caption{Three dimensional plot of the normalized intensity $|\bm{E}_{\perp}\cdot\bm{e}_z|^2/(E_0^2\theta_0^2)$ evaluated at $z=0$ of the $z$-component of the electric field as given by Eq. \eqref{ParField5A}. The intensity pattern shows a structure similar to the $\text{TEM}_{10}$ field distribution of a laser beam, namely, two peaks aligned along the $x$ direction and equally spaced from the origin of the axis $\{x,y\}=\{0,0\}$. This pattern is typical of longitudinal terms.}
\label{figure3}
\end{center}
\end{figure}

\section{Covariance matrix for paraxial Gaussian fields}\label{C}
The Fourier transform of the monochromatic paraxial field 
\begin{align}\label{C10}
\psi(\br,t) =  \frac{ e^{ i k_0 (z - c  t)}}{1+iz/L} \exp \left[ - \frac{1}{w_0^2} \left( \frac{x^2 + y^2}{1+iz/L} \right) \right], 
\end{align}
is calculated by using the standard Gaussian integration formula 
\begin{align}\label{C20}
\int\limits_{-\infty}^\infty  \exp \left( \alpha \, u + \beta u^2 \right) \di u = \left( \frac{\pi}{-\beta}\right)^{1/2} \exp\left(- \frac{\alpha^2}{4 \beta} \right),
\end{align}
where $\alpha$ and $\beta$ are complex numbers with $\text{Re}[ \beta] <0$. 
The key trick is to perform first the integral with respect to $x$ and $y$. In this way a term of the form $1 + iz/L$ is generated and it cancels out with the analogous term in the denominator of the front factor in Eq. \eqref{C10}. The integrals with respect to either $x$ or $y$ read as:
\begin{align}\label{C25}
\int\limits_{-\infty}^\infty & \exp  \left( - i k_u u - \frac{1}{w_0^2} \frac{u^2}{1+i z/L}\right) \di u &  \nonumber \\
& = \pi w_0 (1+i z/L)^{1/2} \exp \left[ - \frac{w_0^2 \, k_u^2}{4}(1+i z/L)\right], 
\end{align}
where $u \in \{x,y\}$ and $w_0^2/4 = L/(2 k_0)$ by definition of $L$. Therefore, after applying twice Eq. \eqref{C25} we obtain
\begin{align}\label{C26}
\int & \exp  \left[ -i \left( k_x x + k_y y \right) \right] \psi(\br,t) \di x \di y&  \nonumber \\
& = e^{ i k_0 (z - c  t)}\pi^2 w_0^2  \exp \left[ - \frac{k_x^2 + k_y^2}{2 k_0/L} (1+i z/L)\right].
\end{align}
 At this point the remaining integral with respect to $z$ is trivial and yields
\begin{align}\label{C30}
\widetilde{\psi}(\bk,t) = & \; e^{-i k_0 c \, t}\frac{(2 \pi)^{1/2}L}{k_0} \exp \left[ - \frac{L}{2 k_0} \left( k_x^2 + k_y^2 \right) \right] \nonumber \\
 & \; \times \delta \left( k_z - k_0 + \frac{k_x^2 + k_y^2}{2 k_0}\right).
\end{align}
 Now, $C_{xx} +C_{yy}$ may be easily calculated by using again the Dirac delta trick outlined in Appendix \ref{A} and by noticing that
\begin{align}\label{C40}
 -k_0 \frac{\partial}{\partial L}\exp &  \left[{ - \frac{L}{ k_0} \left( k_x^2 + k_y^2 \right) } \right] &  \nonumber \\
& =  \left( k_x^2 + k_y^2 \right)\exp\left[ - \frac{L}{ k_0} \left( k_x^2 + k_y^2 \right) \right].&
\end{align}
Thus, in practice, given
\begin{align}\label{C50}
Z_0 = & \; \int e^{ - \frac{L}{ k_0} \left( k_x^2 + k_y^2 \right) }  \left[ \delta \left( k_z - k_0 + \frac{k_x^2 + k_y^2}{2 k_0}\right)\right]^2 \di^3 k \nonumber \\
= & \; g_\epsilon(0) \frac{\pi k_0}{L},
\end{align}
from the definition of $C_{xx} +C_{yy}$ and Eq. \eqref{C40} it follows that
\begin{align}\label{C60}
C_{xx} +C_{yy} =  & \; -k_0 \, \frac{\displaystyle{ 1 }}{Z_0} \frac{\partial Z_0}{\partial L} \nonumber \\
=  & \; \frac{k_0}{L}\nonumber \\
=  & \; \frac{k_0^2\, \theta_0^2}{2}.
\end{align}

\end{document}